\documentclass[prb,twocolumn,showpacs,superscriptaddress,amsmath,amssymb]{revtex4-1}

\usepackage{graphicx}
\usepackage{dcolumn}
\usepackage{bm}

\begin{document}

\title{Study of atomic disorder in Ni-V alloys}

\author{Adane Gebretsadik}
\altaffiliation[Present address: ]{Intel Chandler AZ, USA} 
\author{Ruizhe Wang}
\author{Arwa Alyami}
\author{Hind Adawi}
\author{Jean-Guy Lussier}
\affiliation{Physics Department, Kent State University, Kent OH 44242, USA}
\author{Katharine L. Page}
\affiliation{Spallation Neutron Source, Oak Ridge National Laboratory, Oak Ridge, TN 37831, USA}
\affiliation{Department of Materials Science and Engineering, The University of Tennessee, TN 37996. USA}
\author{Almut Schroeder}
\email{aschroe2@kent.edu}
\affiliation{Physics Department, Kent State University, Kent OH 44242, USA}

\date{\today}

\begin{abstract}
We present a pair distribution function (PDF) analysis from neutron diffraction data of the Ni$_{1-x}$V$_x$ alloy in the Ni-rich regime. Such structural study aims to clarify the origin of the magnetic inhomogeneities associated with the quantum Griffiths phase close to the ferromagnetic-paramagnetic quantum phase transition. 
The PDF analysis successfully reveals the details of the structure and chemical distribution of our Ni$_{1-x}$V$_x$  polycrystalline samples prepared with high-temperature annealing and rapid cooling protocol. 
This study confirms the expectations that all Ni$_{1-x}$V$_x$ samples with $0 \leq x \leq 0.15$ crystallize in a single phase fcc structure with some residual strain. 
The increase of the lattice constant and the atomic displacement parameter with V-concentration $x$ is consistently explained by  
a random occupation of V and Ni-atoms on the lattice, with a radius ratio ($r_{\textrm{V}}/r_{\textrm{Ni}}$) of 1.05. 
Probing alternate, simple models of the local PDF, such as V-clusters or ordered structures (Ni$_8$V, Ni$_3$V) give inferior results compared to a random occupation. This investigation strongly supports that the magnetic clusters in the binary alloy Ni$_{1-x}$V$_x$ originate from Ni-rich regions created from {\it random} occupation rather than from chemical clusters. It reveals that Ni$_{1-x}$V$_x$ is one of the rare examples of a solid solution in a wide concentration regime  (up to $x=0.15$) persisting down to low temperatures ($T=15$\,K).

%
\end{abstract}

\maketitle



\section{\label{sec:intro}Introduction}

Ni-alloys 
remain highly attractive materials for their tunable mechanical and magnetic properties. 
While pure Ni is a very weak ductile metal forming a simple fcc-lattice, 
small amounts of defects typically increase the mechanical strength. Already in a binary alloy
partial substitution of Ni by another d-electron element X allows the formation of partially ordered structures that modify the mechanical properties. 
Ni-superalloys\cite{Kobayashi12} containing local ordered structures within an occupational disordered Ni matrix with defects, are well known for high-temperature applications.
Also, multicomponent alloys \cite{Cantor04} of similar 3d-elements ranging from Cr to Ni, called high entropy alloys \cite{Yeh04}, are promising materials for their mechanical strength. 
The individual local Ni environment and small lattice variations play an essential role. More sophisticated structural methods \cite{Egamibook} beyond the traditional diffraction techniques  are required to resolve local deviations and short-range order \cite{Cowley50}. 

The same is true for the magnetic properties. 
On the one hand, Ni is one of the few elemental ferromagnets (FM) with a high critical temperature $T_c=630$\,K \cite{Boelling68} while on the other hand the magnetism of Ni is very sensitive to changes in the local environment caused by other elements that weaken the magnetic moment and the magnetic order. It is known that
$T_c$ is easily tuned\cite{Bettinelli99} in Ni$_{1-x}$X$_x$ by partial chemical substitution of Ni with another d-element X down to very low values. Ni-alloys seem to provide a good opportunity for  observing magnetic quantum phase transitions (QPT) by reducing the FM ordering temperature $T_c$ down to zero and leaving a paramagnetic phase (PM) without magnetic order. 
Various binary alloys have been studied, which show a suppression of $T_c$ towards 0. Different critical concentrations $x_c$ are extrapolated that depend on the 3d or 4d element X. 
Examples include X=Cr,V with $x_c\approx0.12$, X=Rh,Cu,Pt with $x_c\approx0.4-0.6$ and X=Pt with $x_c\approx0.95$ (see e.g. Refs \onlinecite{amamou73,Rodriguez06}).
 While partial chemical substitution is known to be an effective tuning parameter to apply chemical pressure or for electronic doping to drive through a QPT,  it might modify the critical behavior by introducing {\it disorder} through local atomic and structural inhomogeneities.
The effect of disorder is quite apparent in itinerant ferromagnets:  {\it clean}, i.e. ideally defect-free, homogeneous FM  QPT and {\it disordered} FM QPT are distinctly different \cite{Brando16}.    Already the prominent Ni$_{1-x}$Pd$_x$ \cite{Nicklas99} with only weak disorder does not follow the prediction of a {\it clean} FM \cite{Brando16}. 
Strong disorder might destroy the transition; but under the right circumstances disorder can produce a new exotic quantum critical point\cite{Vojta06} where finite size magnetic clusters play a role. Such novel quantum critical results can only be observed in magnetic alloys which present the proper distribution of random magnetic clusters produced by random defects.  Therefore the full characterization of a disordered magnetic QPT includes a close look at the origin of the disorder. This study probes how {\it ideal  disorder} can be realized in a sample by checking for random static defects.

\begin{figure}
\includegraphics[width=\columnwidth]{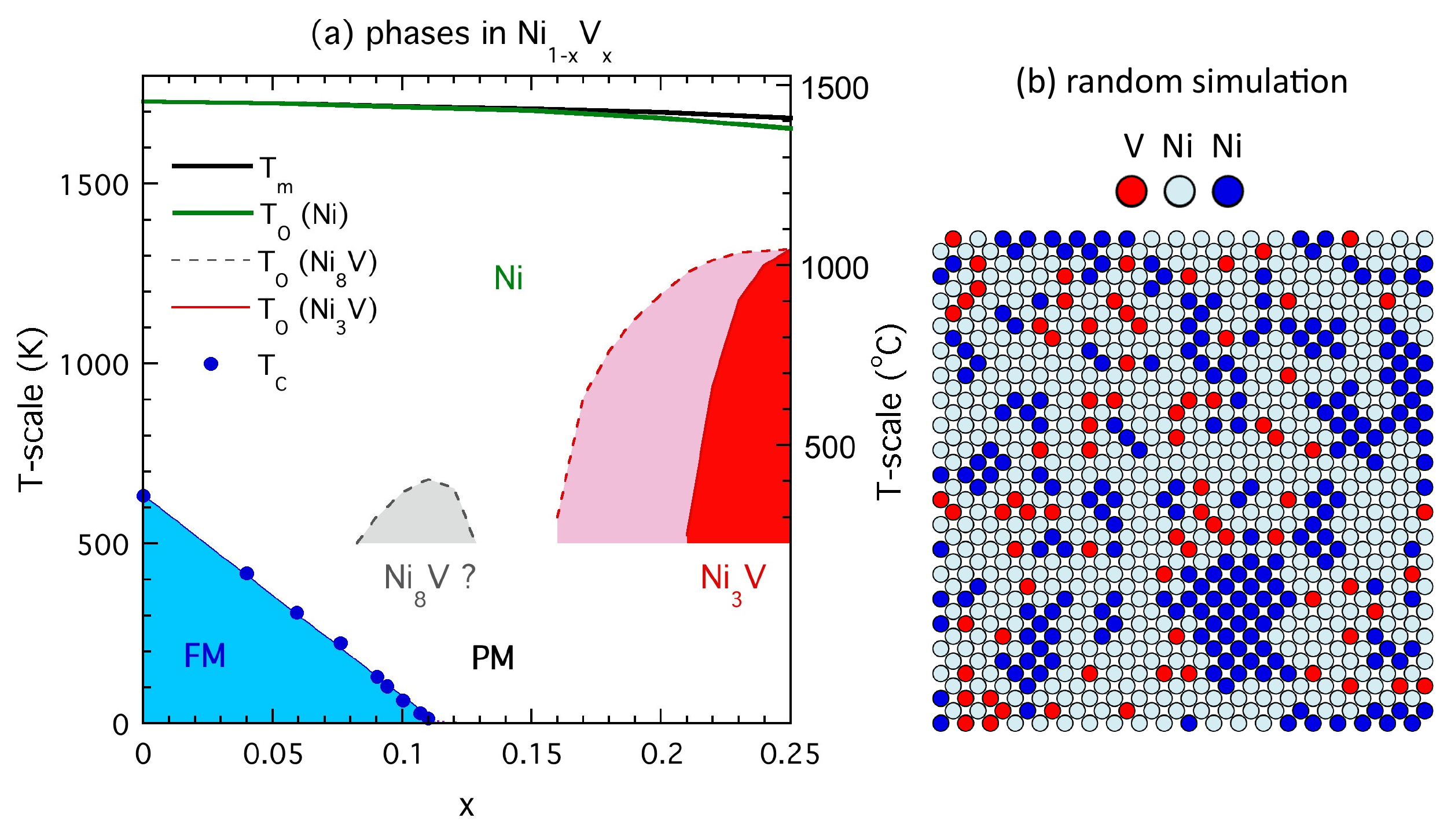}
\caption{(a) Structural  and magnetic phase diagram of Ni$_{1-x}$V$_{x}$ (after Refs \onlinecite{Smith82}, \onlinecite{Wang17}): the melting temperature ($T_m$); the onset of fcc-lattice with random atomic distribution at $T_{\mathrm{o}} $(Ni),  of ordered Ni$_3$V structure at $T_{\mathrm{o}}($Ni$_3$V$)$, of potential Ni$_8$V structure at $T_{\mathrm{o}}($Ni$_8$V$)$, and the magnetic transition at $T_c$ from paramagnetic (PM) to ferromagnetic (FM) phase are shown vs. V-concentration $x$. 
(b) Simulation of random atomic distribution of Ni$_{0.9}$V$_{0.1}$ in xy-plane of fcc lattice: the red circles indicate the random occupation of V. The magnetic response of Ni depends on the neighborhood and is weaker for Ni (in light blue) with adjacent V. The other Ni (in dark blue) mainly contribute to magnetic order or form random magnetic clusters.}
 \label{fig:phase}
\end{figure}

We focus here on the alloy Ni$_{1-x}$V$_x$ which shows indications of a disordered magnetic QPT \cite{Ubaid10,Wang17}. Ni-V with a very small critical concentration $x_c$ also promises the best atomic structure, a solid solution, an fcc lattice with {\it random} atomic occupation. We aim to confirm this with the present study. The local atomic positions of V-atoms are relevant for the magnetic QPT because only the other Ni-atoms seem responsible for the magnetism. 
V differs from the host Ni in the number of 3d-electrons and produces a large magnetic disturbance in  Ni$_{1-x}$V$_x$\cite{CollinsLow65}  by effectively reducing the Ni-moments in its neighborhood. This suppression leads to the rapid average moment reduction \cite{Friedel58} with increasing $x$, up to a small $x_c$,  and to an inhomogeneous magnetization density in Ni$_{1-x}$V$_x$ as illustrated in Fig. \ref{fig:phase}(b). The locations of V are expected to mark the non-magnetic defects that determine the distribution of the remaining {\it magnetic} Ni responsible for the magnetism. These magnetic Ni without any V neighbors contribute to long-range magnetic ordered regions
or to the short-range magnetic clusters 
that lead to the distinct signatures of a disordered QPT\cite{Vojta06,Vojta10}. Magnetization measurements and internal field measurements through $\mu$SR have revealed evidence for such magnetic clusters in Ni$_{1-x}$V$_x$ in both the PM and FM phase\cite{Wang17} close to $x_c$.  
It is a challenge to reveal more details 
of these magnetic clusters, the size distribution and dynamics range are not fully resolved. Recent small angle neutron scattering data give some size estimates\cite{SANS}. However, the ideal prerequisite for random distribution of magnetic clusters are random V occupations. Any other atomic placement would modify the magnetic cluster distribution and the magnetism. Let us first check the prerequisites, how random the V defects are distributed. 

Since any crystalline or chemical defects have a direct impact on magnetic and mechanical properties in Ni-V, 
a thorough structural investigation is essential to reveal  
the quality of the Ni-V samples.
 The phase diagram of Ni-V (Figure\,\ref{fig:phase}(a)) predicts an fcc-lattice with random occupation. However, it is known that the actual chemical structure formation of these binary alloys depends on growth conditions and post-annealing treatments (e.g. Ni-Cu \cite{Tranchita78}). 
 Vanadium clusters or chemically ordered structures change the distribution of Ni-rich regions. The magnetism depends on the immediate Ni-V environment. 
 We chose wide angle neutron diffraction to extract the local pair distribution function (PDF) \cite{Egamibook} of our polycrystalline samples
 to check for deviations from the ideal structure and the ideal random chemical occupations.
This method had been already successful in distinguishing order from occupational disorder \cite{Cu3Au} in a similar binary compound Cu$_3$Au and in recognizing the effect of short-range order. 
The neutron probe offers the advantage of a high contrast between the Ni and V nuclear cross section. (The thermal coherent cross sections are $\sigma$(Ni) $=13.3$ barn, $\sigma(^{58}$Ni$)=26.1$ barn, $\sigma(V)=0.018$ barn). 
Essentially, we are probing the Ni-Ni correlation expecting distinct differences in the PDF peak intensities of the first neighbors for different V occupations. 
Note that this technique does not reveal {\it magnetic} correlations in our FM samples; the magnetic moment contribution is too small to be resolved ($\mu(x_c)\approx 0.02\mu_\textrm{B}$). 
     
 We analyze the atomic pair distribution function (PDF) from a wide angle neutron scattering experiment to probe the local chemical environment in our Ni$_{1-x}$V$_x$ samples (with $x\leq0.15$).
 We will demonstrate that the Ni-V-data are well described by a pure fcc- crystal structure with the expected average Ni environment of a randomly occupied lattice. Comparing the fit quality of different models we can exclude large V-clusters 
 and long-range ordered atomic structures in Ni$_{1-x}$V$_x$ up to $x=0.15$.

\section{\label{sec:exp}Experimental Details}
Polycrystalline spherical samples of Ni$_{1-x}$V$_x$ with V concentrations $x=0$ to 0.15 were prepared by arc melting from
high purity elements (Ni 99.995\%, $^{58}$Ni 99.9\% V 99.8\%), annealed in an evacuated sealed quartz tube at 1000\,$^\circ$C for 3 days, cooled rapidly ($>200^{\circ}$C/min) and investigated by several methods as described in Refs.\
\onlinecite{Ubaid10,Schroederetal14}. The samples with $x=0.110$ and $x=0.123$ were made with the pure isotope $^{58}$Ni and annealed at $1050^{\circ}$C.
Neutron diffraction data of several samples with different V-concentrations $x$ were collected  at the NPDF instrument \cite{NPDF} at the Los Alamos Neutron Science Center.
For this experiment 15-36 pellets with diameter $3-4\,\mathrm{mm}$ were measured for each $x$ inside an aluminum can of diameter 3/8" at 15\,K for 2-3\,h.  
Also,  a powder sample with  $x=0.150$ was investigated at the NOMAD instrument \cite{NOMAD12} at the Spallation Neutron Source (SNS) at the Oak Ridge National Laboratory (ORNL). The powder ($\sim0.3$\,g) was filled inside a glass tube of 2\,mm diameter, and data were collected for 2\,h at 300\,K. 
The NPDF data were reduced with PDFgetN \cite{PDFgetN} (with $Q_{\textrm{max}}=40\,\textit{\AA}^{-1}$ and $Q_{\textrm{min}}=1.8\,\textit{\AA}^{-1}$) to produce the total pair distribution function (PDF) in the form $G(r)$ ready to be modeled with the PDFgui software \cite{PDFgui}. The NOMAD data were reduced and transformed (with $Q_{\textrm{max}}=31.41\,\textit{\AA}^{-1}$) using the automatic data reduction scripts at the NOMAD beamline.

%
%
%
%
%
%
\section{\label{sec:phases} Phase diagram of Ni$_{1-x}$V$_x$}
The binary alloy Ni$_{1-x}$V$_x$ features an apparently simple phase diagram as shown in Figure\,\ref{fig:phase}(a). The ferromagnetic ordering temperature $T_c$ is initially linearly suppressed \cite{Ubaid10} with increasing V concentration $x$, reaching zero toward $x_c=0.116$. Signs of magnetic clusters \cite{Ubaid10,Wang17} are found around $x_c$ in between $x=0.9$ and $x=0.15$. 
Ni-rich Ni$_{1-x}$V$_x$ is expected to crystallize in a simple closed packed cubic fcc-structure as does Ni, below $T_{\mathrm{o}}($Ni$)\approx1400^{\circ}$C. Up to $x=0.15$ no specific chemically ordered structure should form; under the ideal growth condition the elements V and Ni are thought to occupy the fcc lattice sites randomly\cite{Smith82}. 
While it is common for Ni-rich binary alloys to display a random fcc-lattice at high temperatures, an extended perfect solid solution phase down to low temperature is extremely rare and would make Ni-V a remarkable example. 
Typically towards lower temperatures deviations develop that modify strongly the magnetic behavior and magnetic cluster formation. 
E.g. Ni-Pt exhibits a chemically ordered phase \cite{Dahmani85} below $T_{\mathrm{O}}$ or short range order correlations  \cite{Rodriguez06}  if annealed at high temperatures $T_A>T_{\mathrm{O}}$ at the concentration of interest ($x_c\approx0.5$).
No chemically ordered phase is detected in Ni-Cu at $x_c\approx0.5$, but preference for chemical clustering \cite{Vrijen78} is found above a miscibility temperature $T_{misc}$, indicating phase separation at lower temperatures. In both compounds with these different short-range correlations, the onset of magnetism (at $x_c$) depends critically on the chemical structure as the Ni-environment changes with sample preparation \cite{Dahmani85,Tranchita78}. 

No sign of phase separation or any miscibility temperature $T_{\textrm{misc}}$, have been reported for Ni-V.  An ordered structure, Ni$_3$V, is found at higher concentrations below $T_{\mathrm{o}}=1050^{\circ}$C \cite{Smith82}. 
At $x=0.11$ a possible Ni$_8$V structure is indicated in the phase diagram. It only forms if V is substituted with larger elements \cite{Moreen71}, Ta or Nb, below $T_{\mathrm{o}}\approx400^{\circ}$C. It is therefore not expected here as an ordered phase, but short range order (SRO) is possible \cite{Bolloch00}. 
The SRO of selected concentrations $x=1/9,1/4,1/3$ in Ni-V has been studied \cite{Barrachin94,Bolloch00} around $T_{\mathrm{o}}$. The effective pair interaction (EPI) energies were found to be $x$-dependent. While no clustering tendencies of closest neighbors of the same element were noted, ordering tendencies were recognized but get weaker \cite{Bolloch00} towards smaller $x$. 
Before testing for short-range signatures of these alternative structures or potential clustering, we model our data with the pair distribution function (PDF) of the random occupied fcc-lattice.
%
 %
 \section{\label{sec:PDF} Pair distribution function analysis of Ni$_{1-x}$V$_x$}
The pair distribution function (PDF) is essentially the Fourier transform of the total scattering function back into real space to probe for spatial correlations. We assume an ideal isotropic environment (expected in a powder) by averaging over all directions and consider only the modulus of wave vector transfer $Q$ and distance $r$. $G(r)=4\pi r [\rho-\rho_0]$ is a typical form of the PDF used for the PDFgui\cite{PDFgui} software; it gives the contrast between the pair density function $\rho$ from the large distance average $\rho_0$. 
\begin{equation}
G(r)= 2/\pi \int_{Q\textrm{min}}^{Q\textrm{max}} [S(Q)-1]\,Q \sin{(Qr)} \, dQ 
\label{eq:G(r)}
\end{equation}
$S(Q)$ is the normalized total scattering function that includes the Bragg peaks and the diffuse scattering collected at the instrument (after background subtraction and calibration).

\begin{figure}
\includegraphics[width=\columnwidth]{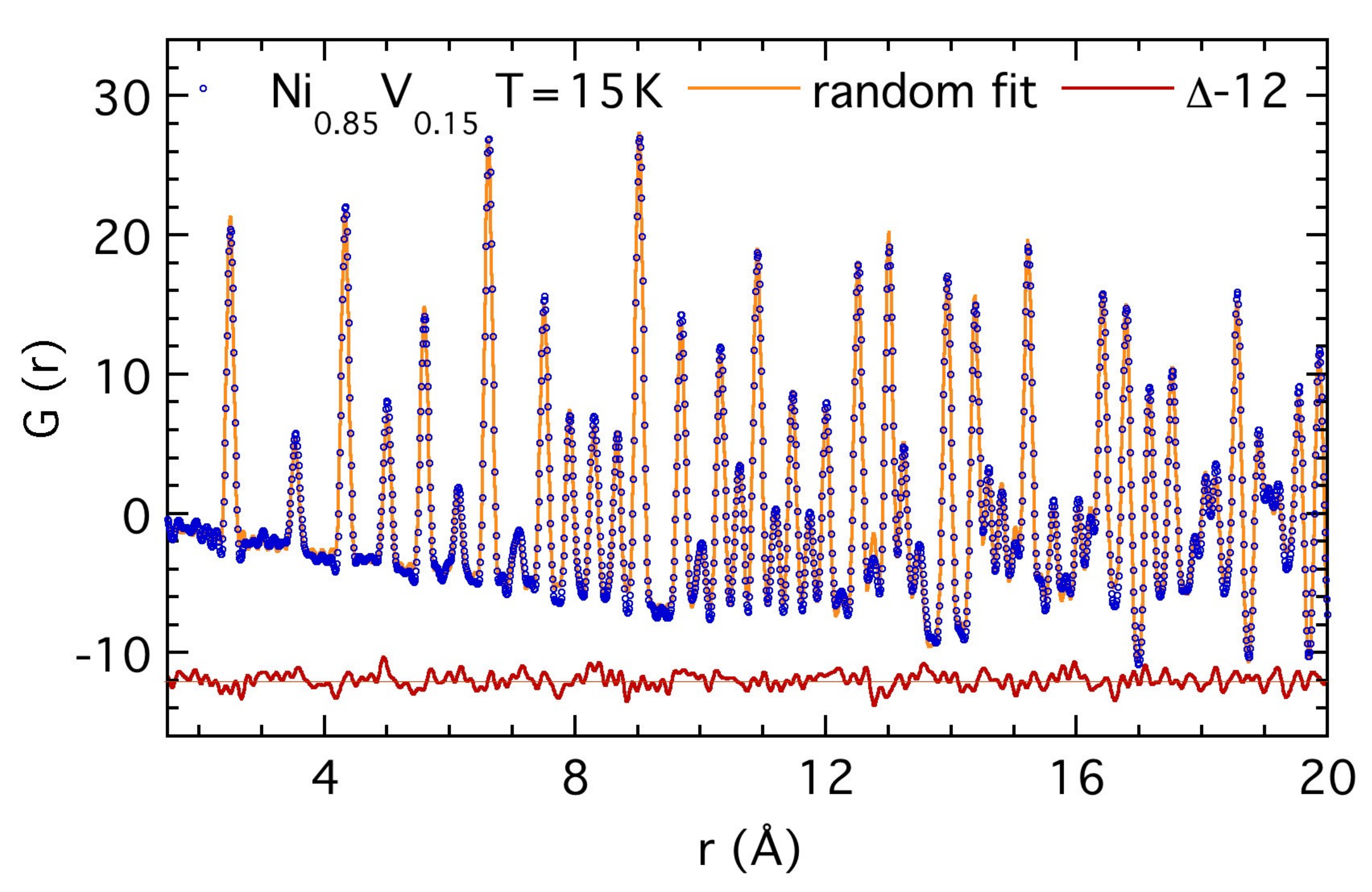}
\caption{Pair distribution function $G(r)$ vs pair distance $r$ of Ni$_{0.85}$V$_{0.15}$ pellet data (blue circles) taken at 15\,K at NPDF with random fit (orange line). The difference, $\Delta$ = data - fit, is shown as a red line shifted by 12 units with residual $Rw=7.9\%$.}
\label{fig:NPDF}
\end{figure}

The typical PDF $G(r)$ of Ni-V is shown in Figure\, \ref{fig:NPDF}. The blue circles mark the $G(r)$ pellet data of Ni$_{0.85}$V$_{0.15}$ taken at low temperatures. All our Ni-V samples with different V concentration $x$, including Ni, produce similar $G(r)$ that looks like the pure Ni powder data \cite{NOMAD12}. 
The Ni-V data are described well with a single phase fcc-lattice with Ni occupation of $(1-x)$ and V occupation of $x$ according to the chemical composition of the sample. This  $random$ fit using a fit range of $1.75\,\textit{\AA}<r <20\,\textit{\AA}$ is shown as an orange line.  The difference $\Delta$ of data and fit is shown underneath shifted by 12 units; the low weighted residual factor $Rw$ is  $7.9\%$. This quick analysis already confirms that the samples do not deviate  much from an fcc lattice with random occupation. In what follows we look more closely to optimize the precision of this statement and reveal more about our samples quality using the PDFgui program \cite{PDFgui}.

With PDFgui we will extract the essential structural parameters, the cubic lattice constant $a$ and the atomic displacement parameter (ADP)  $u$,  at different ranges but also tweak other parameters to optimize the fit quality that gives information about the crystalline quality of the samples.
The ADP is the mean square atomic displacement from equilibrium position of one element averaged over time and sites. 
Since by far the strongest signal comes from Ni we consider only one isotropic parameter  $u=u_{\textrm{Ni}}$ for Ni and chose the same value $u_{\textrm{V}}=u_{\textrm{Ni}}=u$ for V.
The parameter $u$ is extracted from the observed peak width in $G(r)$ that also include effective correlation parameters  and instrumental resolution parameters as explained in Appendix A. 

The $G(r)$ of Ni$_{0.85}$V$_{0.15}$ (in Figure\, \ref{fig:NPDF}) does not present additional peaks besides the fcc-lattice to indicate any secondary phase. The same is true for all Ni$_{1-x}$V$_x$ samples from $x=0$ to $x=0.15$. But the fit quality represented by the weighted residual $Rw$ is not optimal, raising some concerns about structural defects. The peak widths grow larger and the PDF intensity decays faster with distance $r$ than ideally expected for a perfect lattice. We optimize the fit quality systematically by alternatively tweaking control parameters or using two-phase models that model a variation of the lattice parameter $a$ as explained in more detail in Appendix A.  This simple analysis reveals some lattice imperfection, with a lattice variation $\Delta a/a$ of 0.4$\%$ as strain estimate. Since our samples are rapidly cooled after annealing to maintain a random distribution of the V and Ni atoms, such lattice imperfections with strain are expected. This strain is already present in pure Ni and does not increase much  in the V-alloy.

\begin{figure}
\includegraphics[width=\columnwidth]{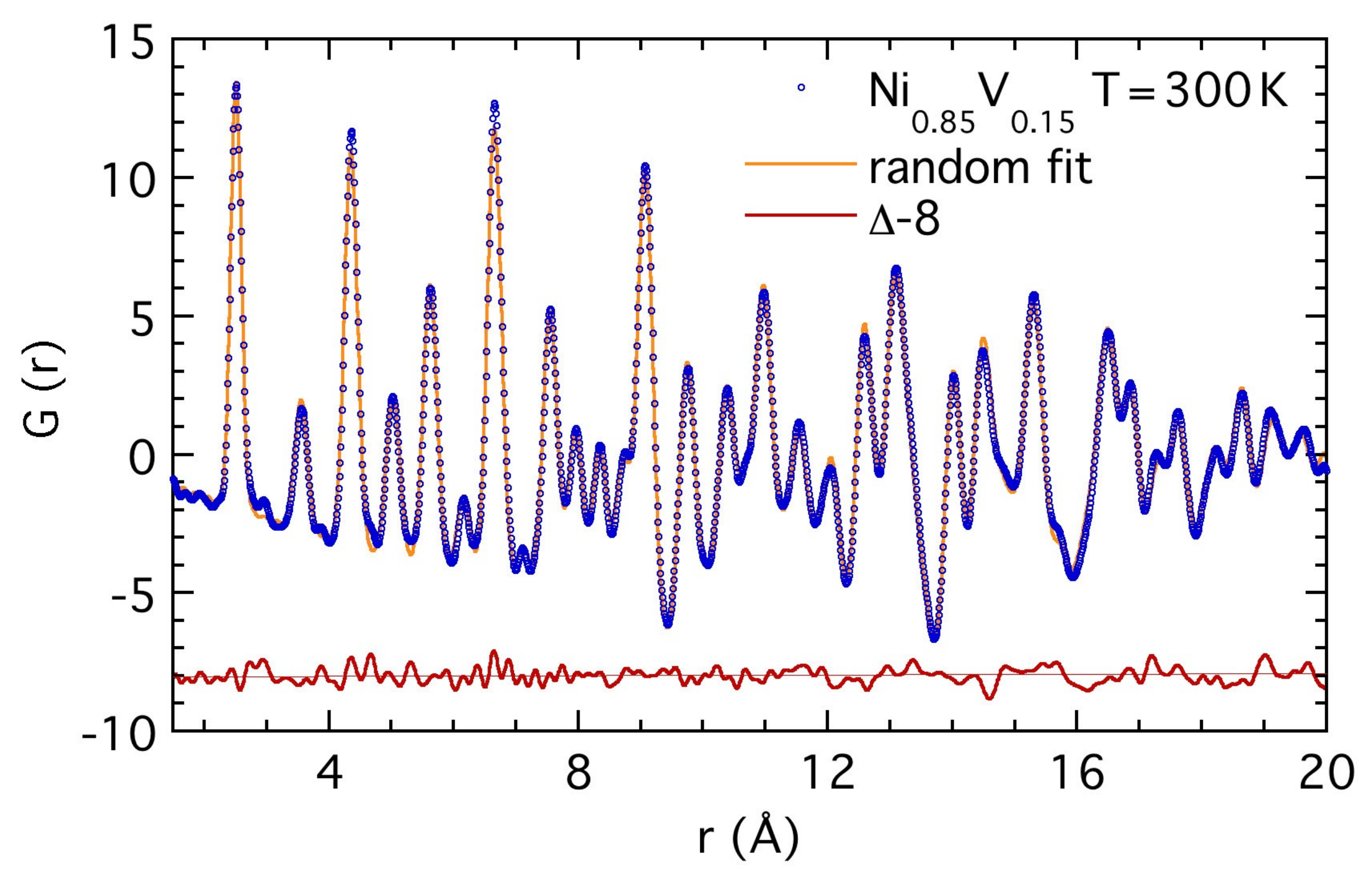}
\caption{PDF of Ni$_{0.85}$V${_{0.15}}$ powder data (blue circles) taken at 300\,K at NOMAD with random fit (orange line). The difference, $\Delta$ = data - fit, is shown as red line shifted by 8 units with residual $Rw=8.45\%$. }
\label{fig:NOMAD}
\end{figure}

The PDF data are noisy with higher $Rw$ from these {\it pellet} samples especially for pure Ni with the largest pellets. To sort out what relates to the internal sample quality or the sample distribution in the can, we performed a confirmation experiment at room temperature on a {\it powder} sample produced from filing down some pellets. Figure \ref{fig:NOMAD} shows the PDF $G(r)$ for the same concentration $x=0.15$ on a powder measured at NOMAD at 300K. The peaks are broader due to thermal motion and the lattice constant is larger because the data were collected at high temperatures, but the fcc-lattice with random occupation describes these powder data as well as the previous pellet data. We notice  that the {\it reduction} of $Rw$ by optimizing control parameters is similar in all samples and is therefore related to lattice imperfections. We conclude that the individual $Rw$ depends rather on the sample distribution. 
In the following study of lattice parameters and models we use the optimized control parameters to account for the strain. All $G(r)$ data up to $r_{\textrm{max}}=20\textit{\AA}$ are presented by fits with optimized control parameters; for short range fits with $r_{\textrm{max}}=7\textit{\AA}$ the original calibrated control parameters are sufficient. The detailed parameters are listed in  Appendix A.\\

%
%
%
%
%
%
%

\begin{figure}
\includegraphics[width=\columnwidth]{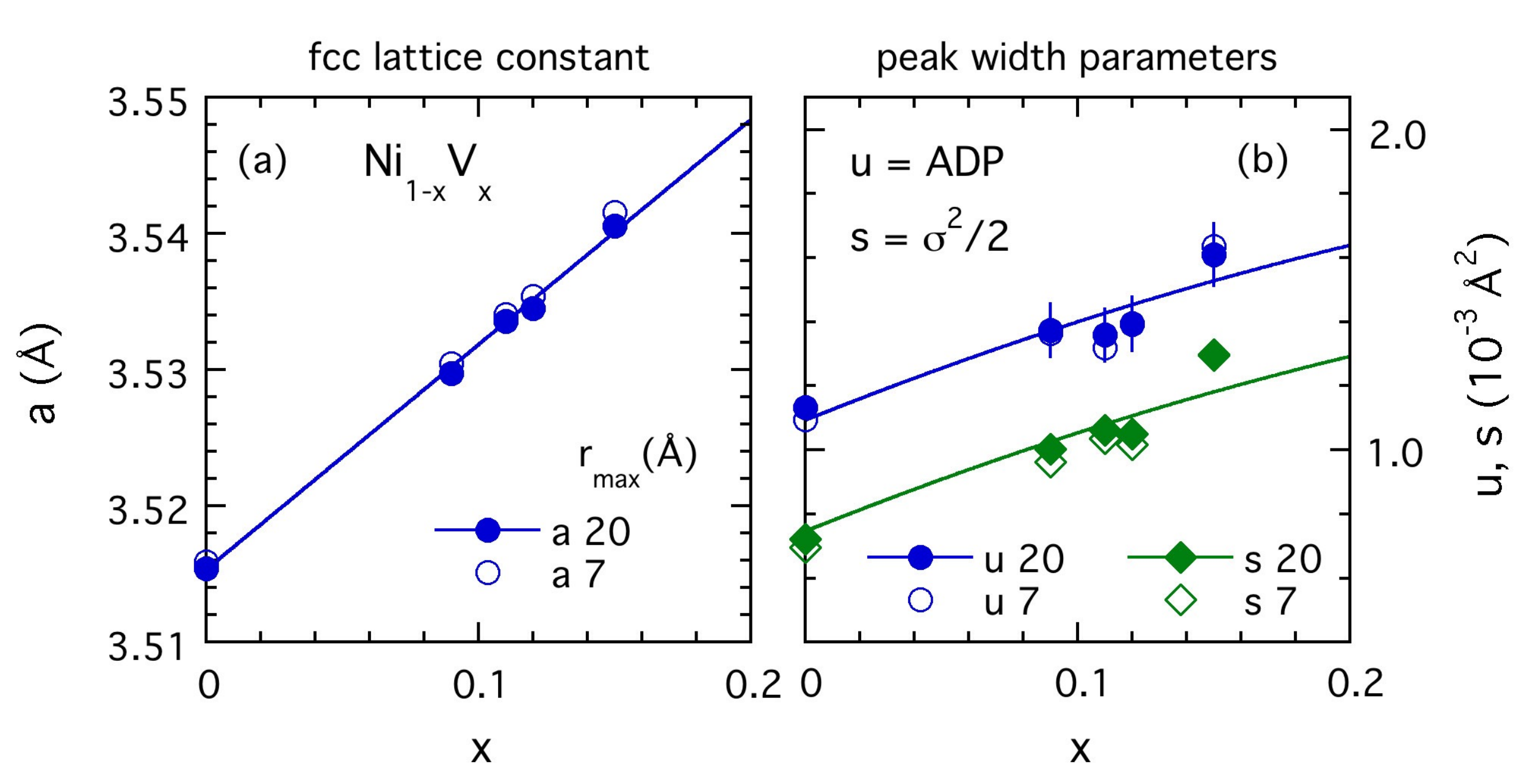}
\caption{(a) Lattice constant $a$, (b) atomic displacement parameter $u$, and half the nearest neighbor distance variance, $s=\frac{1}{2} \sigma^2$, vs. V-concentration $x$ as refined from random fit for Ni$_{1-x}$V$_x$ at $T=15$\,K. Solid (open) symbols show data for $r_{\textrm{max}}=20\,\textit{\AA}$ ($7\,\textit{\AA}$). The line in (a) is a fit of Eq.\,(\ref{eq:ax}) with $a_0=3.5153\,\textit{\AA}$ and $b=0.047$. The lines in (b) follow Eq.\,(\ref{eq:ux}) with the same $a_0$, $b$ and the fit constant $u_0=0.00109\,\textit{\AA}^2$ and $s_0=0.00075\,\textit{\AA}^2$.}
\label{fig:au}
\end{figure}

With meaningful control parameters 
we are ready to evaluate reliably the lattice parameter, $a$ and the ADP $u$, and study their $x$-dependence in Ni$_{1-x}$V$_x$. 
Figure\,\ref{fig:au} presents $a$ and $u$ of all samples at low $T=15$\,K as a function of the V concentration $x$ evaluated for long and short ranges with $r_{\textrm{max}}=20\,\textit{\AA}$ and $7\,\textit{\AA}$. 
The increase of $a$ with $x$ is linear and follows here simply Vegard's law \cite{Vegard21}; the average atomic or ion radius increases with $x$ due to $x$ larger V-atoms with atomic radius $r_{V}$ and $(1-x)$ smaller Ni-atoms with $r_{Ni}$.
\begin{equation}
a(x)=a_0(1+b\,x) \, \, \text{with} \, \, b=(r_\textrm{V}-r_{\textrm{Ni}})/r_{\textrm{Ni}}.
\label{eq:ax} 
\end{equation}
The line in Figure\,\ref{fig:au}(a) is a fit of Eq.\,(\ref{eq:ax}) with $b=0.047$ and $a_0=3.5153\,\textit{\AA}$. Such simple lattice constant increase implies a constant atomic radius ratio of V and Ni (in the fcc-lattice) of $r_\textrm{V}/r_{\textrm{Ni}}=1+b=1.05$. This simple rigid sphere model is also supported by the pure metals. We find the same atomic ratio for V and Ni from the atomic distances at room temperature.   The ratio of the nearest neighbor distances of V (in bcc lattice) and of Ni is  $d_\textrm{V}/d_{\textrm{Ni}}=\sqrt{3/2} \, (3.04\,\textit{\AA})/(3.54\,\textit{\AA})=1.05$ (see e.g. in Ref. \onlinecite{Hull1922}).

The peak width in $G(r)$ also changes with $x$. The extracted ADP $u$ increases with $x$ for $x\leq0.15$ as shown in Figure\,\ref{fig:au}(b). Does this indicate further lattice defects or is it simply related to the different sizes of the V and Ni ions in the lattice? We will predict next the static lattice distortions for the given ion size ratio with random dilution.
The result of the random prediction is shown as solid line in the Figure\,\ref{fig:au}(b). It matches the observed data as discussed in detail below.

We collected data at low $T$, that the APD, $u= u_{\textrm{dyn}}+u_{\textrm{stat}}$, is sensitive to static defects and just includes zero point motion; typically the APB is dominated by thermal motion at high $T$.  
 The observed finite $u_0=0.0011\,\textit{\AA}^2$ of Ni is close to the expected $u_{\textrm{dyn}}$ estimate ($u=0.0013\,\textit{\AA}^2$) using the Debye model \cite{Jeong03} for pure Ni (with low $T$ Debye temperature  $\Theta_\textrm{D}=470$\,K from Ref. \onlinecite{Dixon68}).    
Since only minor variations of $u_{\textrm{dyn}}$ with $x$ are expected due to changes in $\Theta_\textrm{D}$, the main increase of $u(x)$ is caused by $u_{\textrm{stat}}$. The parameter $u$ is determined experimentally from the peak width and the static changes can be estimated by the bond length variances. 

%
%
We predict the bond length variance $\sigma^2$ of a random occupied lattice with mean bond length ($\left<2r\right>\approx2r_{\textrm{Ni}}$) to change with $x$ as $\sigma^2=2x(1-x)b^2r^2_{\textrm{Ni}}$. 
The bond length variance is mainly determined by the atomic displacement parameters $u$ of both neighboring atoms \cite{Jeong03} (see Eq. (\ref{eq:sigma}) in Appendix A).
The ADP shows the same increase with $x$ as half the bond length variance $s=\sigma^2/2$ or half the square of the peak width assuming the experimental resolution is irrelevant and the correlation parameters do not change much with $x$. Expressing the bond length  $2r_{\textrm{Ni}}$ through the fcc- lattice constant $a_0$ of Ni ($a_0/\sqrt{2}=2 r_{\textrm{Ni}}$) leads to a quantitative prediction for the change of $u(x)$  [eq.(\ref{eq:ux})] for random occupation: 
\begin{equation}
u(x)=u_0+\frac{1}{8}b^2a^2_0 \, x(1-x). 
\label{eq:ux}
\end{equation}
The upper solid line in Figure\,\ref{fig:au}(b) shows the expected change of $u(x)$ with the already determined parameters, $a_0$ and $b$, from (a) and $u_0=0.00109\,\textit{\AA}^2$.
We see that the increase is well explained by static defects created only by the given size difference of the atoms with random occupation. 
Also, half of the bond length variance of the closest Ni-neighbor, $s=\frac{1}{2} \sigma^2$,   follows the same fit simply shifted by $0.00034\,\textit{\AA}^2$ (see lower solid line in Figure \ref{fig:au}(b)), confirming that the correlated motion does not change significantly with $x$. The uncertainty of variation in $u_{\textrm{dyn}}$ stemming from the change of effective homogeneous lattice potential upon alloying is assumed to be about $6\%$ as indicated as error bars 
\footnote{The ratio of $\Theta_\textrm{D}($V$)/\Theta_\textrm{D}($Ni$)=380$K/470K (Ref. \onlinecite{Radebaugh66,Dixon68}) or $\Theta_\textrm{D}($Ni$_3$V$)/\Theta_\textrm{D}($Ni$)=520$K/470K (Ref. \onlinecite{Chen14}) might cause a change of  $u_{\textrm{Ni}} \sim1/\Theta_\textrm{D}$ with $x$ of $3\%$ or $6\%$ for $\Delta x=0.15$, if linear interpolated. 
The variations in the low-temperature specific heat coefficient $\beta$ in Ni-V (Ref. \onlinecite{Gregory75}) suggest a possible change in $\Theta_D$ of less then $8\%$ for $x<0.15$.} 
in the values of $u$.
That makes the increase of peak width or $u$ with $x$ consistent with a random occupation as predictable lattice distortion. Thus, we do not see any change of lattice structures or declining crystalline quality evolving with $x$ up to $x=0.15$. 
%

\section{V-Clusters and alternate structure models}
The PDF of our samples is well-described by a fcc-lattice with a random occupation, indicating a solid solution of V and Ni as ideally expected in this concentration range. Possible deviations are the formation of V-clusters, an fcc lattice with locally enhanced vanadium concentration, which lead, in the extreme case, to segregation of large Ni-rich regions from V-rich regions. Another option is a chemical ordered structure with rather alternating Ni and V sites. 
We use here the local PDF to 
test these different Ni-environments 
using the reduced PDF data of Ni$_{0.85}$V$_{0.15}$.  

In a {\it random} fcc-lattice up to a concentration of $x=0.15$, V is expected to have only a few V neighbors out of the 12 nearest neighbors.  
The average V neighbor count is only $z_\textrm{V}=12x=1.8$ for $x=0.15$. Most 
($90\%$) V have less than 4 V-neighbors; 1 or 2 V-neighbors are most likely.
Evidence of V-clusters larger than 4V 
would signal a deviation from the ideal assumptions of a random occupation for $x=0.15$. Since the neutron scattering length for Ni is dominant, the V-V and Ni-V correlations are less obvious in the neutron PDF; the Ni-Ni correlation is the major signal. We expect an average Ni-neighbor count of a Ni-site to be $z=12(1-x)=10.2$ for the random occupied fcc-lattice with $x=0.15$.
Larger V-clusters than expected for random are recognized by larger Ni-rich regions with increased Ni-Ni coordination as directly observed in the first peak intensity in $G(r)$ as discussed below.
 To test for this, we probed different extreme models. 
 
 First, we checked the response of a simulation of a {\it real} random structure for Ni$_{0.85}$V$_{0.15}$, a finite supercell phase with $5^3$ cubic fcc unit cells where 
500 Ni/V were placed once with the probability of $0.85/0.15$.
This {\it random cell} containing some small V clusters produces similar fit results to the previous {\it random model} that just used one unit cell with the same fractional occupation. The fit quality $Rw$ is similar, as recorded in Table \ref{tab:modelpara}.

%
\begin{table}
\caption{\label{tab:modelpara} Fit quality of different models describing the local pair correlations in Ni$_{0.85}$V$_{0.15}$ for the two experimental data sets at 15\,K and 300\,K. The weighted residual factor $Rw$ is listed for different fit ranges $r_{\textrm{max}}$. The concentration of the second phase (*) is about 15\% determined with $r_{\textrm{max}}\approx 7\,\textit{\AA}$ (see text for details).}
\begin{ruledtabular}
\begin{tabular}{lllll}
& 15\,K  &&  300\,K  \\
\hline
$r_{\textrm{max}} $ & $6.9\textit{\AA}$ & ($20\textit{\AA}$) & $7.1\textit{\AA}$ & ($20\textit{\AA}$) \\
\hline
model: & $Rw(\%)$ &  $Rw(\%)$ & $Rw(\%)$& $Rw(\%)$  \\ 
\hline
fcc-random & 7.87 & (7.94) &  6.93 & (8.45) \\
random cell 5a & 7.92 & (8.04) & 7.01 & (8.66) \\
\hline
V4-cluster & 8.77 & (8.68) &   7.94 & (10.1) \\
V13-cluster & 8.55 & (8.77) & 7.72 & (9.55) \\
V38-cluster & 8.87 & (8.92) & 8.22 &  (10.3)\\
\hline
Ni$_8$V & 8.23 & (8.49) &  7.96 & (9.61) \\
Ni$_3$V  & 10.2 &  (17.7) &12.3 & (21.4)\\
\hline
random+Ni$_8$V* & 7.88 & (8.99) &  6.15 & (8.09) \\
random+Ni$_3$V* & 7.35 & (8.60) & 5.93 & (8.51) \\
random+Ni* & 7.74 & (12.7) & 6.70 & (10.9) \\
\end{tabular}
\end{ruledtabular} 
\end{table}

\begin{figure}
\includegraphics[width=\columnwidth]{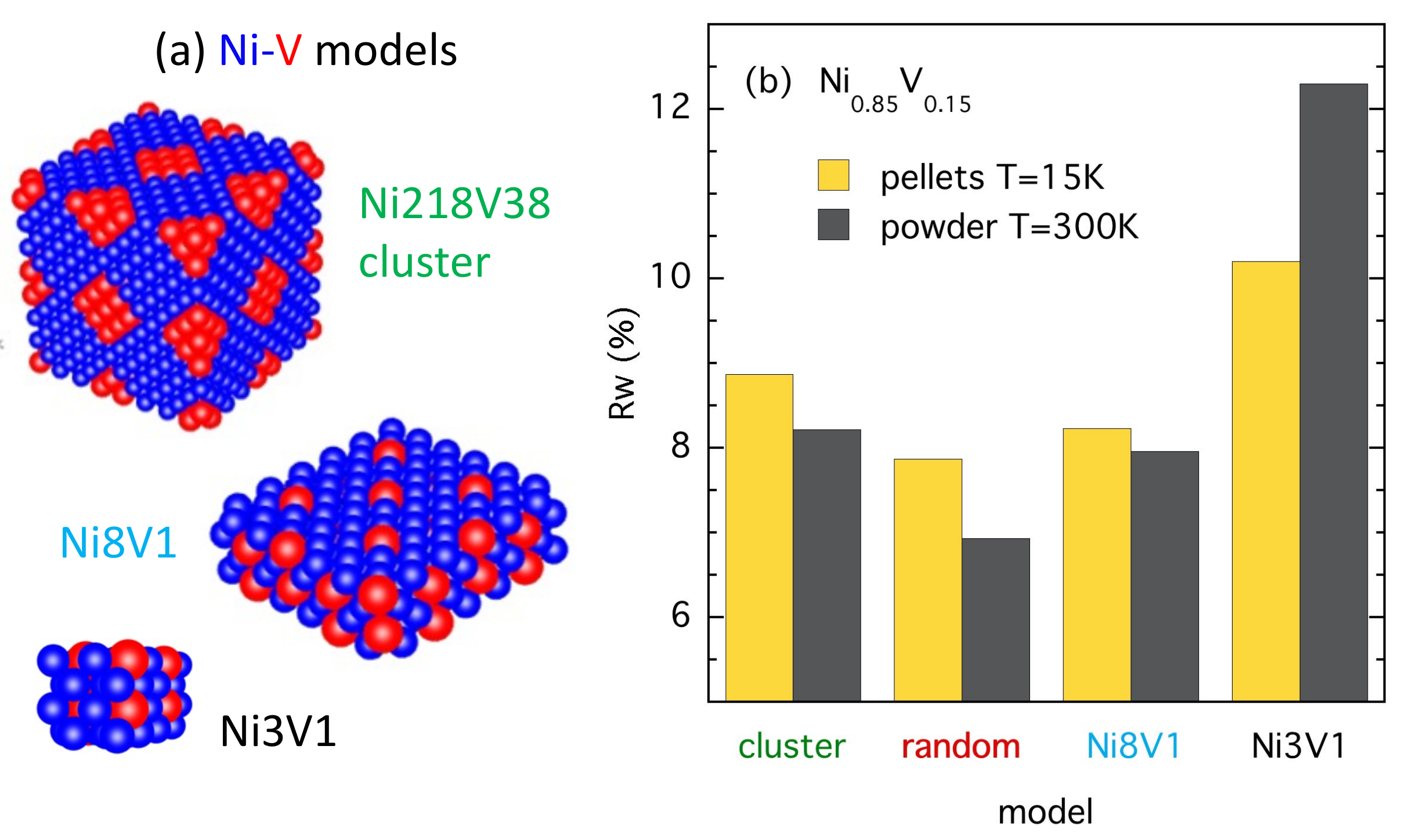}
\caption{ (a) View of different alternate models of Ni-V with red V atoms and blue Ni atoms: V38 cluster model and structures are displayed with $2\times2\times2$ unit cells. (b) Fit quality of different models for pellet and powder data shows the random model with lowest residual factor $Rw$.}
\label{fig:RWmodel}
\end{figure}

\begin{figure}
\includegraphics[width=\columnwidth]{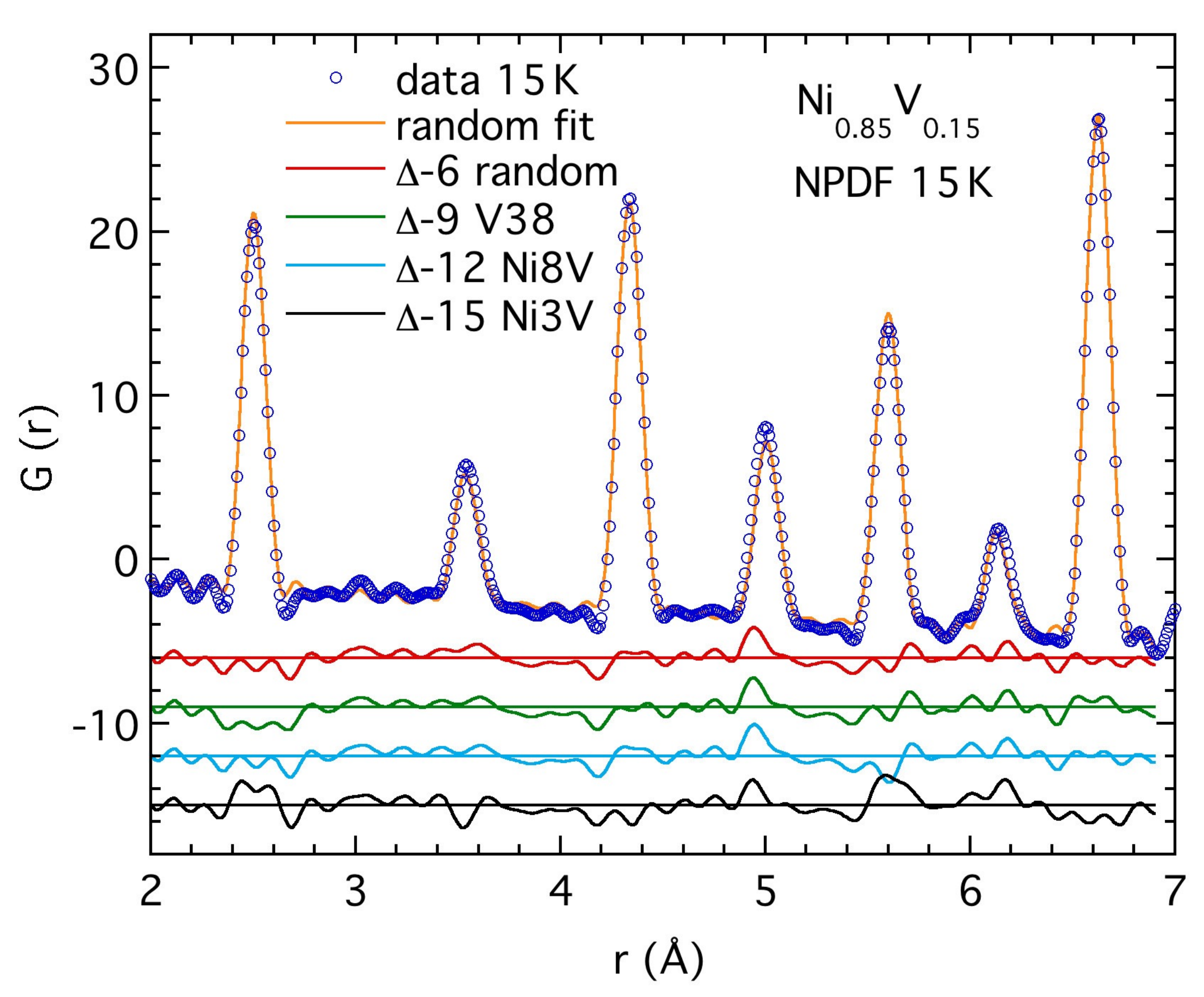}
\caption{Comparison of different models of local pair distribution, Ni$_{0.85}$V$_{0.15}$ data (taken at 15\,K at NPDF) and difference ($\Delta$=data-model); models are fcc-lattice with random occupation, V-clusters, Ni$_8$V and Ni$_3$V structure.}
\label{fig:NPDFmodel}
\end{figure}

To probe larger vanadium clusters we constructed a supercell phase in PDFgui that contained large V-clusters far away from each other: 38 V were placed in a spherical arrangement on an enlarged fcc-lattice ($4\times4\times4$ cubic unit cells) with 256 atoms. This V38 model is illustrated in Figure \ref{fig:RWmodel}. The V-clusters are $\sim 7\,\textit{\AA}$ in size and are placed $\sim 14\,\textit{\AA}$ from center to center, so that the distance in between (from edge to edge) is $\sim 7\,\textit{\AA}$. This simple model of a V-cluster with a single size reflects about the proper concentration of the sample.  Analyzing the short-range correlation with restricted $r_{\textrm{max}}\approx7\,\textit{\AA}$ below the distance between the clusters allows testing the effect of a V-cluster without including the cluster-cluster correlation introduced in this periodic model.   
Figure \ref{fig:NPDFmodel} presents $G(r)$ of both models, the random fit and the V38 fit, with reduced $r_{\textrm{max}}=6.9\,\textit{\AA}$ using the 15\,K NPDF data for Ni$_{0.85}$V$_{0.15}$. 
The V38 model does not create a dramatic change in peak intensity. However, a distinct change is noticed in the difference $\Delta$ between data and fit, in particular at the first peak in $G(r)$ at $\sim2.5 \textit{\AA}$. This nearest neighbor peak intensity is sensitive to the average Ni-Ni coordination z. $\Delta$ of V38 presented in Fig. \ref{fig:NPDFmodel} by the green line (shifted by 9 units) shows more deviations than $\Delta$ of the random model shown as red line above (shifted by 6 units). The V-cluster model with higher Ni-Ni first neighbor coordination $z$ than the random model does not improve the fit. The better fit with reduced $Rw$ value and lower $z$ remains the random fit compared to the V38-cluster fit (see in Figure\,\ref{fig:RWmodel}(b) and Table\,\ref{tab:modelpara}). 
The refined parameters for the random distribution are listed in Appendix A in Table \ref{tab:15fitpara}, and any deviations for other models are found in Appendix B.

For the study of smaller V-clusters we had to make compromises of reduced concentration and reduced edge to edge distance.
We prepared V4-clusters  by defining 4 V within 32 atoms in a $2\times2\times2$ fcc-supercell. The V-concentration of the model ($x=0.125$) is a bit lower than the sample concentration $x=0.15$. With a cluster size of $\sim2.5\textit{\AA}$ and a closest distance between the centers of $\sim7\,\textit{\AA}$, the edge to edge distance is rather short $\sim5\,\textit{\AA}$, but still allows  probing mainly a single cluster correlation with $r_{\textrm{max}}$ of $\sim7\,\textit{\AA}$. Also, a V13-cluster was prepared by defining 13 V within 108 atoms in a $3\times3\times3$ supercell. The V-concentration of this model is $x=0.12$. With a cluster size of $\sim 5\,\textit{\AA}$ and a distance between centers $\sim11\,\textit{\AA}$, the edge to edge distance is $\sim6\,\textit{\AA}$. These smaller V-cluster models produce a similar  PDF as the V38-cluster model with the same characteristic large first peak (not shown). The $Rw$ factors are all higher than the value of the random model (see Table\,\ref{tab:modelpara}).

%
\begin{figure}
\includegraphics[width=\columnwidth]{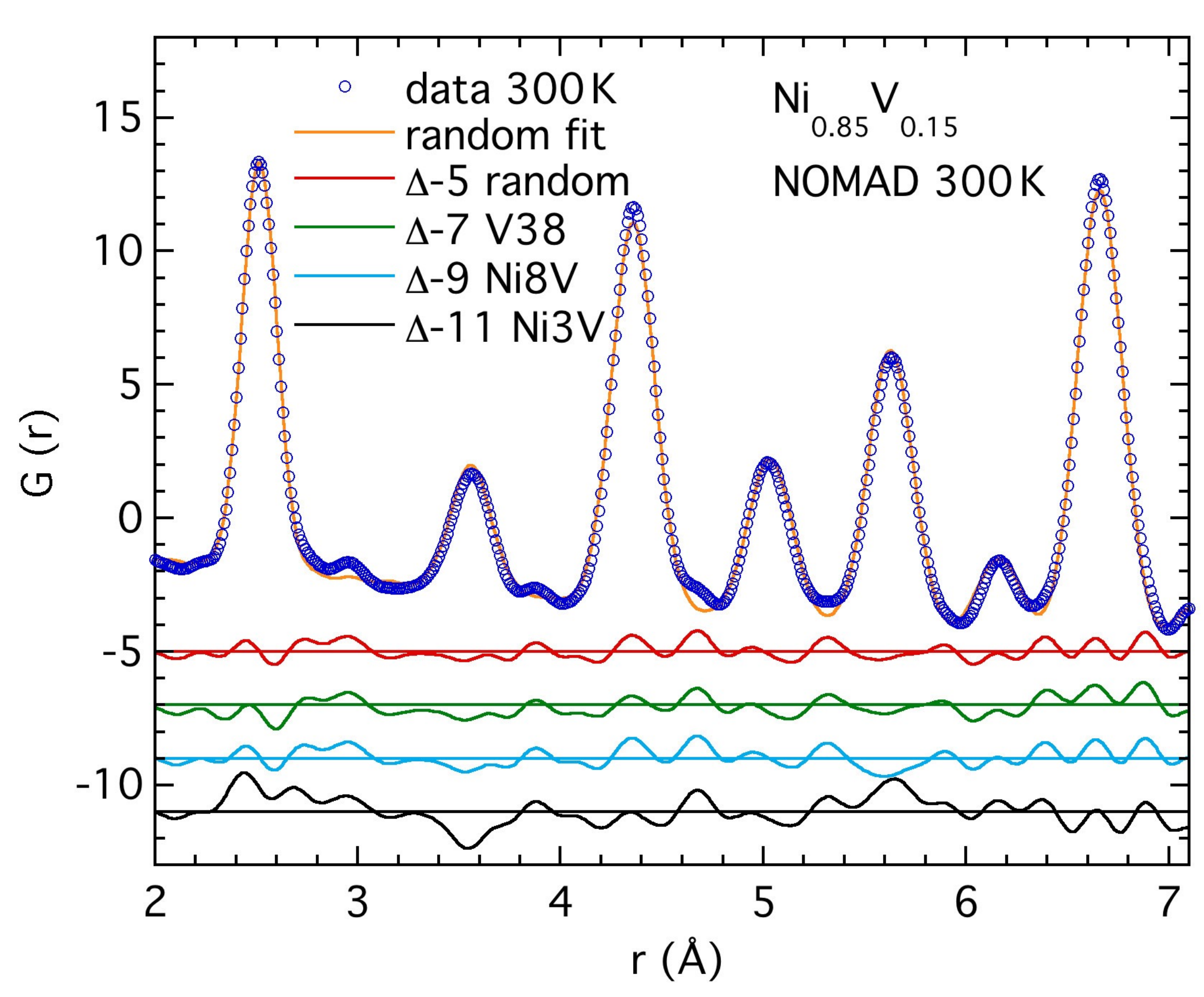}
\caption{Comparison of same models of local pair distribution as in Fig. \ref{fig:NPDFmodel}. The data are Ni$_{0.85}$V$_{0.15}$ powder data taken at 300\,K at NOMAD.}
\label{fig:NOMADmodel}
\end{figure}

The same V-cluster models were applied to the powder data of Ni$_{0.85}$V$_{0.15}$ collected at NOMAD at 300\,K. Figure\,\ref{fig:NOMADmodel} displays the PDF fit results for the random and the V38-cluster model together with the data, in the same order as Figure\,\ref{fig:NPDFmodel} presents the 15\,K data. 
The residuals $Rw$ of the different V-cluster fits as shown in Table\,\ref{tab:modelpara} are all consistently larger than the $Rw$ of the random fit. Table\,\ref{tab:15fitpara} in Appendix A lists the refined parameters with the instrumental parameters for the random model. Most parameters remain similar for the other V-cluster models;
Appendix B comments on some minor deviations. 
Although the fit quality of these different models does not change much, 
the local PDF provides a clear distinction between the models.  The intensity of the first peak matches well the Ni-Ni coordination of a fcc-lattice with random occupation and clearly deviates from the increased (Ni-Ni) neighbor count of the V-cluster models. It does not provide any evidence  for  large V-clusters in Ni-V. 

Other deviations from random occupied fcc-lattice are chemical ordered superstructures in a binary alloy. We are probing here short-range and long-range order of Ni$_8$V and Ni$_3$V (see models in Figure\,\ref{fig:RWmodel}). 
The first potential chemical ordered structure in the Ni-rich region is Ni$_8$V. 
Ni$_8$Nb and Ni$_8$Ta order in this ``Ni$_8$Nb"-structure since the radius ratio is sufficiently large \cite{Moreen71} ($r_{\textrm{Nb}}/r_{\textrm{Ni}}\approx r_{\textrm{Ta}}/r_{\textrm{Ni}}\approx1.14$). 
The Ni$_8$Nb-structure is a body-centered tetragonal structure (space group I4/mmm) with 9/2 fcc unit cells with $a=b=\sqrt{9/2} c$. The Nb/V-site has no similar neighbors; Ni has 1 or 2 Nb/V neighbors; the average Ni-Ni first neighbor count is $z=10.5$. Although this ordered Ni$_8$V structure does not form as long-range ordered phase at $x=0.111$ short-range correlation can still be relevant in Ni$_{1-x}$V$_x$ in a wider concentration range \cite{Bolloch00}.

%
%
The Ni$_8$V structure was prepared (as a $3\times3\times1$ fcc-supercell with V at the origin and face center) to model the PDF data of Ni$_{0.85}$V$_{0.15}$ for short distances with $r_{\textrm{max}}\approx7\,\textit{\AA}$ with PDFgui under the same condition as the random fit. See model in Figure\,\ref{fig:RWmodel}. The difference $\Delta$ of data-model is displayed in Figure\,\ref{fig:NPDFmodel} and Fig.\,\ref{fig:NOMADmodel} as a third line (in blue) shifted down by some units. The residual $Rw$, recorded in Table\,\ref{tab:modelpara}, is small, but still larger than the random value. The difference between $z=10.2$ and 10.5, the effective average Ni-Ni neighbor count is not very large, so farther neighbor correlations become relevant for the formation of the ordered structure.
%
%

The other superstructure Ni$_3$V forms from the disordered fcc phase below $T_o=1045^{\circ}$C in a higher concentration range around $x=0.25$, depending on sample growth conditions and heat treatments \cite{Smith82,Moreen71}. Ni$_3$V crystallizes in the SO$_{22}$ structure, a body-centered tetragonal structure with 2 fcc unit cells along the $c$-direction \cite{Lin92}, where $c/a>2$.  V has only Ni neighbors, and Ni has 4 V neighbors that the Ni-Ni coordination $z=8$ is very small.
The short-range correlation of this Ni$_3$V structure were tested (using $1\times1\times2$ fcc supercell with V at origin and body center) as shown in Fig.\,\ref{fig:RWmodel}. The difference $\Delta$ of data and model is presented as lowest (black) line in Fig.\,\ref{fig:NPDFmodel} and Fig.\,\ref{fig:NOMADmodel}. It shows obvious deviations, e.g. at the first peak in $G(r)$, and leads to the largest residual $Rw$ as listed in Table\,\ref{tab:modelpara}. This organized structure reduces the Ni-Ni correlation to $z=8$ which contradicts the experimental data. The refined parameters (see more in Appendix B) are similar to the random results, only the lattice constants differ with $c/a=2.006$. 

We presented a detailed analysis of the most V rich sample Ni$_{0.85}$V$_{0.15}$, which would be expected to be the most prone to V clustering.  Figure\,\ref{fig:RWmodel}(b) summarizes the $Rw$ for the main models clearly identifying the random model as the best description with the lowest $Rw$. Applying the same analysis on the other Ni-V samples with $x<0.15$ gives similar results. The random model remains the best description for all. For all $x$, the residual factor $Rw$ increases consistently by $\sim0.5-1\%$ modeling the local PDF with the V-cluster model compared to the random model. 
The  $x=0.110$ pellet sample is described equally well with the Ni$_8$V structure or with the random model. For short ranges  ($r_{\textrm{max}}=6.9\,\textit{\AA}$), the Ni$_8$V model is slightly better than the random fit [$Rw($Ni$_8$V$)=14.4\%<Rw(\mathrm{random})=14.7\%$]; for larger ranges ($r_{\textrm{max}}=20\,\textit{\AA}$) the Ni$_8$V fit becomes somewhat worse [$Rw($Ni$_8$V$)=14.9\%>Rw(\mathrm{random})=14.8\%$]. These samples were annealed at high $T$ for random distribution. To what extent short-range order of the Ni$_8$V remains in these samples cannot be resolved because of insufficient statistics. 
These data already demonstrate that no obvious V-clustering and no large-scale phase separation occurs in Ni-V. 

\section{Two-Phase models}
Besides testing alternate models such as random or super structure, these two models and their contribution can be probed simultaneously for the same data set in a two-phase (2P) model. This is a simple way to notice deviations from a random occupied fcc lattice in regions within the sample and recognize atomic short range correlations.  
Modeling the PDF of the Ni$_{0.85}$V$_{0.15}$ data with the random phase and the ordered phase Ni$_3$V (with a contribution of about $15\%\pm 8\%$) leads to a better description with a reduced $Rw$ than the pure random model if the fit regime is restricted to $r_{\textrm{max}}\approx7\,\textit{\AA}$. 
Expanding the fit regime to $r_{\textrm{max}}=20\,\textit{\AA}$ does not improve $Rw$ compared to the pure random model, signaling that only  {\it short}-range correlations of Ni$_3$V are present.  The refined lattice parameters 
are consistent with the expected values (see Appendix B). 
When the ordered phase is replaced by a second random phase (with independent lattice constant and contribution) the fit quality declines. This confirms that already weak short range correlation of the Ni$_3$V structure are present in Ni$_{0.85}$V$_{0.15}$.
We also tested for short range correlation of Ni$_8$V with a 2P model. 
 The best fit yield a  Ni$_8$V phase contribution of about $14\% \pm7 \%$ with reasonable lattice parameters (see Appendix B). The residual does not show much improvement, the Ni$_8$V local environment is not very distinct from the random Ni-environment. More distinct longer range correlations different from random are not confirmed as expected for our samples annealed at high temperatures.

A 2P model can also be used to probe phase separation of Ni and V or large concentration gradients in Ni-V by separating pure Ni regions from diluted Ni-V regions. Phase separation of Ni+Ni-X has been suggested for Ni-Rh \cite{Teeriniemi15} forming below a miscibility temperature from a disordered fcc phase at higher temperatures. Modeling the PDF data with a pure Ni-phase (with constrained Ni parameters) and a random occupied Ni$_{1-x}$V$_x$ phase ($x\geq0.15$) with adjustable parameters yields a slightly better fit than the single-phase random model if the fit range is restricted to a short range of $r_{\textrm{max}}\approx7\,\textit{\AA}$. Table\,\ref{tab:modelpara} shows the $Rw$ factors. The indicated contribution of the Ni phase is $12\%\pm5\%$. Probing for a Ni-rich region up to $r_{\textrm{max}}=20\,\textit{\AA}$ returns only zero or a negative contribution. If $12\%$ of the pure Ni phase is imposed, the $Rw$ factor increases. Therefore, large Ni regions beyond the random statistics
can be excluded. 
Pure Ni regions become more likely within a smaller volume of radius $r_{\textrm{max}}$.  
The observed value of $\sim12\%$ is still higher than the probability of a pure $x=0$ region within $r_{\textrm{max}}=7\,\textit{\AA}$ ($<1\%$) but matches the probability of $x=0$ below $3\,\textit{\AA}$ (of $12\%$ of Ni with only Ni neighbors)  in a randomly diluted NiV sample with $x=0.15$.   
Some small size Ni-rich regions are noticed in $x=0.15$ that point to minor local deviations from the average concentration and the ideal crystal structure. 
It was shown that in Ni-Cr nanoparticles \cite{Bohra15} with a diameter of $d<10$ nm, different chemical environments were found at the surface compared to the bulk due to Cr segregation to the surface.  We expect here much larger crystallites in our polycrystalline samples, but different Ni-environment at grain boundaries are certainly possible. Grain boundary-aided nucleation of growth of the ordered Ni$_3$V structure was investigated in a melt spun Ni$_{0.75}$V$_{0.25}$ alloy \cite{Singh05}.  
These PDF results support that in Ni-V, minor local concentration gradients are present but no phase separation on larger scales. 

\section{Conclusion}
We present a detailed pair distribution (PDF) analysis from neutron scattering data of the Ni$_{1-x}$V$_x$ alloy. 
This study answers the main question by confirming that our Ni-V samples form indeed a solid solution at low temperatures. The results also demonstrate that the local PDF is a powerful method to probe the relevant Ni-environment in the Ni$_{1-x}$V$_x$ samples to reveal many details. No secondary ordered phase besides the fcc lattice is found. The fcc lattice is the best model when V and Ni are occupying the fcc lattice sites at random. V-cluster models are worse descriptions. The results show no distinct phase segregation of V and Ni-rich regions. Other chemically-ordered structure models show deviations from the data due to the different local environment. The PDF analysis reveals at most weak short-range correlations of Ni$_3$V 
in Ni$_{0.85}$V$_{0.15}$. 
Also, the increase of the lattice constant and the atomic displacement parameters (ADP) with $x$ is consistent with the simple packing of solid spheres (of V and Ni-atoms) in a fcc lattice with occupational disorder. 
This PDF analysis concludes that Ni$_{1-x}$V$_x$ is a system with potential short-range order at specific concentrations, but not prone to chemical clustering like Ni-Cu. 
Ni$_{1-x}$V$_x$ shows more preference for ordering than for clustering. V-clustering as a cause for magnetic cluster formation can be excluded.  
The chemical ordering correlations are rather weak for $x\leq0.15$. 
The local Ni-environment in Ni$_8$V is not very distinct from the random occupation.
That makes Ni$_{1-x}$V$_x$ a remarkable system that favors random occupation when prepared with high annealing temperatures and cooled down rapidly. 
 
How much any weak remnants of chemical ordering impacts the magnetism here could be studied further with optimized samples and more advanced models.    
Regions with short range order (SRO) might not lead to the formation of Ni-rich regions but could still modify the magnetic cluster distribution. Deviations from {\it perfect random} are most likely to occur close to $x\approx0.11$, the concentration of the ordered phase that is close to the critical concentration $x_c=0.116$, where the ferromagnetic order breaks down with most dominant {\it magnetic} clusters.   
The analysis of $x=0.110$ or better of $x=0.111$ could be improved using powdered samples to resolve more subtle differences between ordered and disordered models (with overall reduced $Rw$). 
The PDF of samples prepared with different annealing temperatures $T_A$  showing different SRO can reveal the impact on magnetic clusters through comparison with magnetic measurements. 
To directly observe the {\it magnetic} clusters in these polycrystals the mPDF method\cite{Frandsen2014,Frandsen2022} (extracting the magnetic PDF) remains too challenging in this small moment system but direct neutron scattering measurements like SANS seem feasible to characterize magnetic correlation at $x\approx 0.11$\cite{Schroeder2020,SANS}.\\


\section{Acknowledgement}
This work has benefitted from the use of the NPDF instrument at the Los Alamos Neutron Science Center, Los Alamos National Laboratory, funded by the US Department of Energy.  Part of this research was conducted at the NOMAD instrument at the Spallation Neutron Source, a US Department of Energy Office of Science User Facility operated by Oak Ridge National Laboratory.


%
%
%
\appendix

\section{PDF parameters of random model}

These sections list the detailed fit parameters from the PDF analysis.
We used the PDFgui program \cite{PDFgui} to analyze the data. Besides the crystal lattice and atomic position parameters it also determines their variations, the atomic displacement parameters ADP and contains two simple control parameters to adjust to the resolution of the instrument configuration. 
 The ADP parameter $u$ is extracted from the observed peak width in $G(r)$.  The effective parameters $\delta_2$ and $\delta_1$ are introduced to correct the uncorrelated width for a correlated motion of close pairs \cite{Jeong03}. 
The square of the peak width (HWHM), the experimental variance of the mean bond length $\sigma^2$ is then produced by these simple fit parameters in the PDFgui  program:
\begin{equation}
\sigma^2=2u (1-\delta_2/r^2 -\delta_1(T)/r + (r\,Q_{\textrm{broad}})^2)
\label{eq:sigma}
\end{equation} 
%
 $Q_{\textrm{broad}}$ and $Q_{\textrm{damp}}$ are the instrumental control parameters that model the instrumental peak width and effective intensity decay of $G(r)$ in PDFgui. The values of both parameters are usually determined by a Si-powder experiment and are called $Q_0$ here for both instruments.   
For low temperatures (15K) $\delta_1=0$, for high temperature (300K) $\delta_1$ dominates that we kept $\delta_2=0$.

Table \ref{tab:xfitpara} displays the refined fit parameters and compares the $Rw$ for different setups for some selected Ni$_{1-x}$V$_x$ samples. 
Aiming to {\it optimize} the fit quality by varying parameters we can characterize the crystalline properties of the sample. 
The fit quality of the random fcc-lattice with the calibrated instrumental parameter setting (called $Q_0$) is satisfactory, yielding a residual $Rw=11\%$ for $x=0.15$, but  improves to $Rw=8\%$ by increasing the parameters to larger values (called $Q_L$).  E.g. the $Q_0$ values for NPDF are $Q_{\textrm{damp}}=0.006\textit{\AA}^{-1}$ and $Q_{\textrm{broad}}=0.002\textit{\AA}^{-1}$, the $Q_L$ values  are $Q_{\textrm{damp}}=0.02\textit{\AA}^{-1}$ and $Q_{\textrm{broad}}=0.035\textit{\AA}^{-1}$. This implies that the peak widths grow larger and the PDF intensity decays faster with $r$ than ideally expected for a perfect lattice. More details of modeling resolution effects and their impact on data analysis can be found in Ref. \onlinecite{Olds2018}. 
While the total $Rw(x)$ for all samples varies with $x$ rather non-systematically (from about 20\% - 10\%) instead of  increasing, $Rw$ is always reduced by the same amount of $2-3\%$ when optimizing the control parameters from $Q_0$ to $Q_L$. 
%
\begin{table}
\caption{\label{tab:xfitpara} Refined fit parameters of Ni$_{1-x}$V$_x$  at 15\,K  of the fcc-random model: fcc-lattice constant $a$, (isotropic) atomic displacement parameter $u$, correlated motion parameters $\delta_2$, $\delta_1$, weighted residual factor $Rw$ within fit range $r_{\textrm{max}}=20\,\textit{\AA}$ using optimized instrumental parameters $Q_L$   ($Q_{\textrm{damp}}=0.02\,\textit{\AA}^{-1}, Q_{\textrm{broad}}=0.035\,\textit{\AA}^{-1}$). For comparison $Rw$ of alternative fits are shown with calibrated $Q_0$ parameters ($Q_{\textrm{damp}}=0.006\,\textit{\AA}^{-1}, Q_{\textrm{broad}}=0.002\,\textit{\AA}^{-1}$) employing single-phase and two-phase (2P) model with lattice constant $a_2>a_1$, where $\Delta a/a=2(a_2-a_1)/(a_2+a_1)$. *$Rw$ with nanoparticle size {\it dia} $\geq 300\,\textit{\AA}$.  }
\begin{ruledtabular}
\begin{tabular}{lllll}
Ni$_{1-x}$V$_x$ & x=0 &x=0.9 & x=0.15 \\
\hline
$a (\textit{\AA})$ &  3.51540(1) & 3.52970(1) & 3.54058(1)\\
$u (\textit{\AA}^2)$ & 0.001132(2) & 0.001373(3) &0.001610(3) \\
$\delta_2 (\textit{\AA}^2)$ &  2.32(4) & 1.73(4) & 1.26(3) \\
$Rw(\%): [Q_L]$ & 17.2 & 14.0 & 7.94 \\
\hline
$Rw(\%): [Q_0]$ & 18.8 & 15.7 & 11.0 \\
\hline
$Rw(\%): [Q_0] \, 2P$ & 17.0/17.0* & 13.8/13.8* & 8.26/7.96* \\
$\Delta a/a(\%)$ & 0.353 & 0.365 & 0.378 \\
\end{tabular}
\end{ruledtabular}
\end{table}
\begin{table}
\caption{\label{tab:15fitpara} Control and refined fit parameters of Ni$_{0.85}$V$_{0.15}$ for the two data sets at 15\,K and 300\,K of the fcc-random model. Note the similar fit quality ($Rw$) and parameters for different settings: for $r_{\textrm{max}}\approx 7\,\textit{\AA}$ with $Q_0$ and for $r_{\textrm{max}}=20\,\textit{\AA}$ with $Q_L$ or with $Q_0$ using 2P model, see also Table \ref{tab:xfitpara}.
}
\begin{ruledtabular}
\begin{tabular}{llllll}
Ni$_{0.85}$V$_{0.15}$ & 15 K & & 300 K & \\
$r_{\textrm{max}} (\textit{\AA})$ & 20 & 6.9 &  20 & 7.1 \\
\hline
$a_1 (\textit{\AA})$ & 3.53388(3) & 3.54158(6) &  3.5608(3) & 3.5615(8) \\
$a_2 (\textit{\AA})$ & 3.54728(3)  \\
$u (\textit{\AA}^2)$ & 0.001618(5) & 0.001635(8) &   0.0077(1)& 0.0079(3) \\
$\delta_2 (\textit{\AA}^2)$ & 1.58(4) & 1.30(4) &  &  \\
$\delta_1 (\textit{\AA})$ & & &  1.37(5)&1.37(7) \\
\hline 
$Q_{\textrm{damp}} (\textit{\AA}^{-1})$ & 0.006 & 0.006 &  0.033 & 0.018\\
$Q_{\textrm{broad}} (\textit{\AA}^{-1})$ & 0.002 & 0.002 &  0.040 & 0.019\\
\hline
$Rw(\%)$ & 8.26 & 7.87 &  8.45 & 6.93\\
\end{tabular}
\end{ruledtabular}
\end{table}
%

Note that especially  the largest $Rw$ is observed for $x=0$, which cannot relate to the sample quality but to the sample arrangement. Our pure Ni samples contained pellets with the largest size of $\sim4$mm that lead to the most inhomogeneous distribution within the sample can. We suspect that the sample density variation of several pellets instead of the ideal isotropic powder is responsible for additional wiggles in the PDF data that causes the high $Rw$. 
For comparison, we performed another experiment on a powder sample produced from filing down some pellets for $x=0.15$.
The proper instrumental parameter parameters with low $Q_0$ values produce a residual factor of $Rw=11.5\%$ that reduces to $Rw=8.45\%$ with enhanced $Q_L$ values as shown in Figure \ref{fig:NOMAD}. The optimized experimental parameter and other refined parameters are listed in Table \ref{tab:15fitpara} for the two ranges of interest. This fit quality compares well to similar PDF model of NOMAD data of 2\,g of pure (not annealed) Ni powder reported with $Rw=6.2\%$ \cite{NOMAD12}.

The common {\it reduction} of $Rw$ for all $x$ by increasing the instrumental parameters from $Q_0$ to $Q_L$ at different instruments relates most likely to the sample quality.
Alternatively, these lattice imperfection can be modeled by simple tools in PDFgui through extra parameters, keeping the calibrated values $Q_0$. Through a two-phase (2P) model with two fcc-lattices that only differ in lattice constants ($a_2>a_1$) the strain $\Delta a/a$ can be estimated where $\Delta a = (a_2-a_1)$, and $a= (a_2+a_1)/2$. Also a nanoparticle diameter $dia$ is available in PDFgui to estimate a finite crystallite size. 
The best 2P model fit yield a lattice variation or effective strain $\Delta a/a$ of the order of $0.3-0.4\%$ for all $x$ as shown in Table\, \ref{tab:xfitpara}. This value is consistent with the alternate description using an increased $Q_{\textrm{broad}}=Q_L$ value expecting $\Delta a/a \approx 2 \sqrt{u}\, Q_{\textrm{broad}}$.  The increased $Q_{\textrm{damp}}=Q_L$ value  corresponds to a finite crystallite size in the order of $100\textit{\AA}$  ($dia \approx 2/Q_{\textrm{damp}}$). Within the 2P model the fit quality improves only a bit for $x=0.15$ from $Rw=8.26\%$ to $Rw=7.96\%$ with a finite $dia\approx300\,\textit{\AA}$, while for $x<0.15$ $Rw$ remains unchanged with insignificant high $dia>300\,\textit{\AA}$ values.  


Since both alternative models, two-phase (2P) with $Q_0$ or one phase with $Q_L$ yield essentially the same fit with similar parameters and fit qualities $Rw$ we use further the $Q_L$ setup for this range of $r_{\textrm{max}}=20\,\textit{\AA}$. 
These results also remain similar for a reduced fit range ($r_{\textrm{max}}\approx 7\,\textit{\AA}$) with proper resolution $Q_0$ as expected (see Table \ref{tab:15fitpara}), the main setup we use to probe alternative atomic models.

\section{PDF parameters of other 1P and 2P models}
Here we discuss the essential fit parameters for testing other models than random using single phase and two phases within PDFgui.  
The PDF of Ni-V with $x=0.15$ is fitted with the same control parameters for the short range of $r_{\textrm{min}} = 1.75\textit{\AA}$ to $r_{\textrm{max}}\approx7\textit{\AA}$ for all models: from models with random occupation, to cluster models with selected V-cluster sizes to superstructures (Ni3V, Ni8V). 
Table \ref{tab:modelpara} shows the fit quality through the weighted residual factor $Rw$. The detailed fit parameters are not very different from the random model as listed in Table \ref{tab:15fitpara}. For full information we comment on some deviating parameters of the other models. We call $a_{15}$, $u_{15}$, $\delta_{15}$ the values for the random model.
The cluster models produce the same parameters except the correlation parameter is smaller, decreasing further with increasing the clusters size to V38, where $\delta_1$ becomes $1.17$ and $\delta_2$ reduces to $\delta_2=0.49$. Also the refined parameters of the Ni8V structure are similar to the random fit with $c=a_{15}$.
The Ni3V structure allows two different lattice constants with $a=0.999a_{15}$ and $c/a=2.006$. The ADP is somewhat larger with $u=0.0021 \textit{\AA}^2$ (15K), $u=0.0087 \textit{\AA}^2$ (300K).

We probed 2 phase models to check for short-range order. The most obvious ordered phase is Ni3V. $G(r)$ of Ni$_{0.85}$V$_{0.15}$ is fitted with $r_{\textrm{max}}\approx7 \textit{\AA}$ using 2 phases, one with random occupation and one with the ordered phase Ni3V. The best fit yield a contribution of $15\%\pm8\%$ of Ni3V with a reduced $Rw$ as listed in Table \ref{tab:15fitpara}. The refined lattice parameters are $a_{\textrm{random}}=1.001a_{15}$ and $a_{31}=0.993a_{15}$ with $c_{31}/a_{31}=2.03(1)$ for the random and the ordered Ni3V-phase, respectively.  Using Ni8V for the second ordered phase yields a similar contribution of $14\%\pm7\%$ for the best fit but not a distinct improvement in $Rw$ compared to the pure random model. The refined parameters are  
 $a_{\textrm{random}}=1.001a_{15}$ and $c_{81}=0.998a_{15}$ for the random and the ordered Ni8V phase, respectively.
  
%
%

%

%
\end{document}